\documentclass[11pt]{llncs}

\usepackage{fullpage}
\usepackage{amsmath,amssymb}
\usepackage{times}
\usepackage{graphicx,color}
\usepackage{multirow}
\usepackage{subfigure}
\usepackage{setspace}
\usepackage{mathpartir}
\usepackage{comment}
\usepackage{fancyvrb}
\usepackage{xspace}
\usepackage{todonotes}
\definecolor{darkblue}{rgb}{0.0,0.0,0.6}
\definecolor{darkgreen}{rgb}{0.0,0.6,0.0}
\definecolor{darkred}{rgb}{0.6,0.0,0.0}

\definecolor{background}{HTML}{EEEEEE}

\usepackage{listings}
\newcommand{\cci}[1]{\texttt{#1}}

\newcommand{\Data}{\mathrm{Data}}
\newcommand{\func}{\mathit{func}}
\newcommand{\port}{\mathit{port}}

\newcommand{\guard}{\mathit{guard}}
\newcommand{\source}{\mathit{src}}

\newcommand{\ignore}[1]{}
\newcommand{\ie}{i.e.,}

\newcommand{\true}{\ensuremath{\mathtt{true}}\xspace}
\newcommand{\false}{\ensuremath{\mathtt{false}}\xspace}
\newcommand{\goesto}[1][]{\stackrel{#1}{\longrightarrow}} 

\newcommand{\valu}[1]{\mathbf{#1}}

\newcommand{\Pm}{\ensuremath{\mathcal{S}}\xspace}         

\newcommand{\caig}{\ensuremath{\mathcal{OLP}}\xspace}
\newcommand{\biptool}{\ensuremath{\mathcal{B}ip\mathcal{SV}}}

\newcommand{\var}{\ensuremath{\varphi}}

\newcommand{\be}{\begin{itemize}}
\newcommand{\ee}{\end{itemize}}
\newcommand{\bdn}{\begin{description}}
\newcommand{\edn}{\end{description}}
\newcommand{\bn}{\begin{enumerate}}
\newcommand{\en}{\end{enumerate}}

{\end{list}}

\begin{document}
\sloppy
\graphicspath{{figures/}}

\title{From High-Level Modeling Towards Efficient and Trustworthy Circuits}
\author{}

\institute{}

\maketitle

\begin{abstract}
Behavior-Interaction-Priority (BIP) is a layered embedded 
system design and verification framework that provides 
separation of 
functionality, synchronization, and priority concerns to 
simplify system design and to establish correctness by 
construction. 
The framework comes with a runtime engine and a suite
of verification tools that uses D-Finder and NuSMV as 
model checkers. 
In this paper we provide a method and a supporting tool
that takes a BIP system and a set of invariants
and computes a reduced sequential circuit with a 
system-specific scheduler and with a designated output that 
is \true when the invariants hold. 
Our method uses ABC, a sequential circuit synthesis and 
verification framework to 
(1) generate an efficient FPGA implementation of the system, 
and to (2) verify the system and debug it in case a 
counterexample was found. 
Moreover we generate a concurrent C implementation of the circuit that
can be directly used as a simulator. 
We evaluated our method with two large systems and our results
outperform those possible with existing techniques. 
\end{abstract}

\section{Introduction}
\label{sect-intro}

\begin{figure}
\resizebox{.9\columnwidth}{!}{
  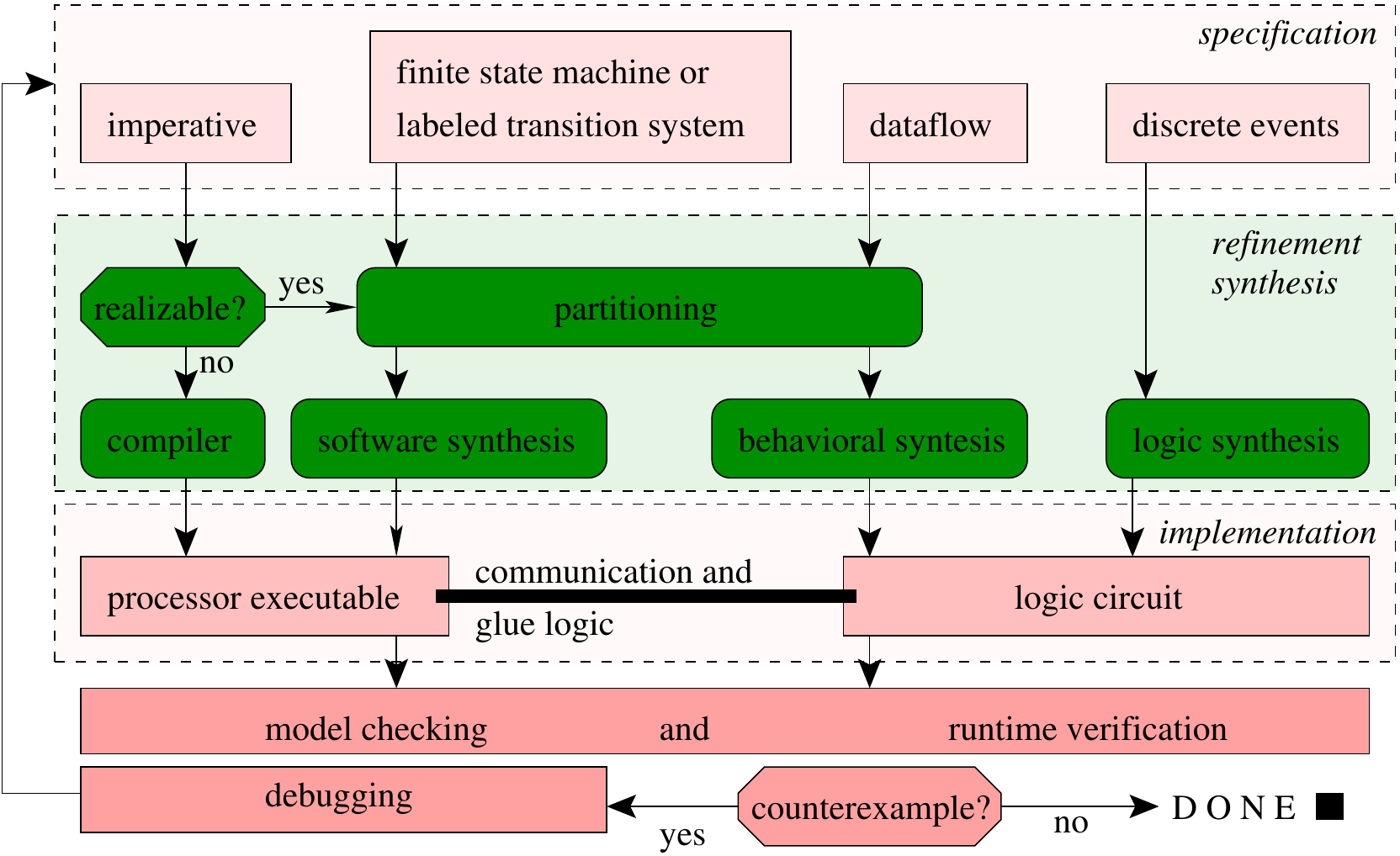
}
\caption{Embedded system specification, refinement, and implementation stages}
\label{fig:flow}
\end{figure}

In recent years, {\em embedded systems} have witnessed a large 
expansion, especially with  the emergence of automotive 
electronics, mobile and control devices.
An embedded system is a composition of {\em heterogeneous}
intellectual property (IP) components.
Figure~\ref{fig:flow} shows a typical flow of the composition process where the
components are specified as imperative programs, finite state machines (FSM), labeled 
transition systems (LTS), data flow networks, and discrete event based circuits. 
Computations in embedded systems are subject to several 
physical and architectural 
constraints that render the separation between software and 
hardware design impractical~\cite{henzinger2006embedded}.
The partitioning task, often done manually, decides whether a component is to 
be implemented as a programmed process or as a realtime logic circuit. 
A plethora of software, behavioral, and logic compilation and synthesis techniques are
used in the process~\cite{metropolis2}.

The Behavior-Interaction-Priority (BIP) framework 
is a {\em Component-Based System} (CBS) design framework that uses a dedicated 
language and tool-set to support a rigorous and layered design flow for embedded 
systems.  
BIP allows to build complex systems by coordinating the behavior of a set of 
atomic components~\cite{bip11}.
BIP makes use of (1) the DFinder~\cite{dfinder} compositional  
and incremental verification tool-set, and (2) the NuSMV~\cite{nusmv} model checker, 
to check the correctness of BIP systems. 
However, DFinder \cite{BBL14} does not handle data transfer between components, 
and it does not support checking for invariants other than deadlock freedom. 
Additionally, for complex systems, NuSMV often suffers from the state space explosion 
problem~\cite{sipser2006introduction}, and fails to perform its verification tasks.

ABC~\cite{brayton2010abc} is a transformation-based 
verification (TBV)~\cite{KuBa01} framework that operates on 
And-Inverter Graphs (AIG); semi-canonical Boolean netlists with
memory elements, and iteratively and synergistically 
employs powerful reduction, abstraction and decision algorithms such as 
retiming~\cite{KuBa01}, 
redundancy removal~\cite{HmBPK05,KuMP01,BjesseC00,aziz-fmsd-00}, 
logic rewriting~\cite{BjBo04}, interpolation~\cite{McMillan03}, 
and localization~\cite{Wang03}, 
symbolic model checking, bounded model checking, induction, 
interpolation, circuit SAT solving, 
and target enlargement~\cite{MoGS00,MoMZ01,HoSH00,BaKuAb02,Hari05expert}.

In this paper, we present a method and a supporting tool (\biptool)
for embedded system synthesis, runtime verification,
and model checking with a cycle based execution model.
The method leverages transformation-based synthesis and verification techniques 
as follows. 

\begin{enumerate}
\item The method takes a BIP system and a set of invariants and generates 
  an AIG circuit with an output therein that is $\mathit{true}$ iff the system 
  is deadlock free, and satisfies the system invariants. 
  The method passes the generated AIG circuit to ABC for verification. 
  ABC either proves correctness or produces a counter example where the 
  system violates an invariant. 
  This enabled us to find defects and prove systems that were not 
  possible using DFinder and NuSMV. 
\item  The supporting tool \biptool~ provides a debugging mechanism where the 
  counter example is mapped back to the original BIP system. 
  The debugging tool is integrated with a wave form visualization tool \cite{bybell2010gtkwave}.  
\item The method generates an FPGA implementation of the BIP system with a 
  system-specific execution framework. 
  The FPGA implementation is passed to ABC synthesis reduction algorithms 
  which reduce the area and the critical time of the FPGA implementation 
  by removing latches and logic gates. 
  To the best our knowledge, we are the first to synthesize a BIP system directly 
  into an FPGA. 
\item The method generates a concurrent C implementation that simulates the BIP 
  system with a system-specific execution framework. 
\end{enumerate}

BIP uses a runtime engine to simulate its execution semantic. 
The main loop of the engine consists of the following steps:
\begin{enumerate}
\item Each atomic component sends to the engine its current location.
\item The engine enumerates the list of interactions in the system, 
  selects the enabled ones based on the current location of the atomic 
  components and eliminates the ones with low priority.
\item The engine non-deterministically selects an interaction out of the enabled interactions.
\item Finally, the engine notifies the corresponding components and schedule their transitions for execution. 
\end{enumerate}
We differ in that, the system specific scheduler is a bit vector of interactions directly embedded in the implementation. 
The interaction bit vector evaluates in real-time and directly depends on the locations and the values of the variables of the input system. 
The system specific execution framework empirically reduces the space and time requirements for the C simulation and the FPGA execution. 

Several frameworks for the design and verification of embedded systems exist. 
We briefly introduce them here and discuss and compare to them later in 
Section~\ref{sec:related}.
Metropolis~\cite{metropolis1,metropolis2} is a design framework that
takes a Metropolis Meta Model (MMM) description of an embedded system 
and generates a SystemC~\cite{systemc} based simulator of the system.
It has also a path to SIS~\cite{brayton92sis} which is a synthesis predecessor tool 
of ABC, and a path to SPIN for model checking~\cite{HolzSpin97}. 
SystemC~\cite{systemc} in turn is a design framework based on C++ that allows
system components to communicate through ports, interfaces, and channels.
Extensions to SystemC such as ForSyDe~\cite{SanderJ04} restrict the 
expressiveness to enable formal verification tools to handle the system. 
In brief, our method supports the synthesis, model checking, and runtime verification 
concerns of embedded systems using tool independent semantics across the three concerns
by embedding the execution model of the embedded system in the generated systems 
for each concern. 
This allows for simple debugging and design flow cycle iterations. Furthermore, 
the use of AIG circuits for synthesis and model checking allows our method to leverage
the mature and rich literature of logic synthesis techniques. 

The remainder of this paper is organized as follows. In Section \ref{sec:bip}, we recall the necessary concepts of the BIP framework. Section \ref{sec:this} defines one loop program (\caig). Section \ref{sec:sequential} formalizes sequential circuit and shows how to translate a sequential circuit into \caig. Section \ref{sec:bip2aig} shows how to translate a BIP system into \caig. Section \ref{sec:implem}
describes \biptool{}, a full implementation of our framework and some benchmarks. Section \ref{sec:related} discusses related work. Section \ref{sec:conclusion} draws some conclusions and perspectives.

\section{BIP - Behavior Interaction Priority}
\label{sec:bip}
%
We recall the necessary concepts of the BIP framework~\cite{bip11}.
BIP allows to construct systems by superposing three layers of desin: Behavior, Interaction, and Priority.
The \emph{behavior} layer consists of a set of atomic components represented by transition systems. 
The \emph{interaction} layer provides the collaboration between components. 
Interactions are described using sets of ports. 
The \emph{priority} layer is used to specify scheduling policies applied to the interaction layer, given by a strict partial order on interactions.
%
\subsection{Component-based Construction}
%
BIP offers primitives and constructs for designing and composing complex behaviors from atomic components. Atomic components are Labeled Transition Systems (LTS) extended with C functions and data. Transitions are labeled with sets of communication ports. 
Composite components are obtained from atomic components by specifying interactions and priorities.
\subsubsection{Atomic Components.}
An atomic component is endowed with a finite set of local variables $X$ taking values in a domain $\mathit{\Data}$. Atomic components synchronize and exchange data with each others through \emph{ports}.
\begin{definition}[Port]
A port $p[x_p]$, where $x_p\subseteq X$, is defined by a port identifier $p$ and some data variables in a set $x_p$ (referred to as the support set). We denote by $p.X$ the set of variables assigned to the port $p$, that is, $x_p$.
\end{definition}
\begin{definition}[Atomic component]
An atomic component $B$ is defined as a tuple $(P,L,$ $T,X,\{g_\tau\}_{\tau \in T}, \{f_{ \tau}\}_{\tau \in T})$, 
where:
\begin{itemize}
\item $(P,L,T)$ is an LTS over a set of ports $P$. $L\ignore{=\{l_1,l_2,\ldots,l_k\}}$ is a set of control locations and $T \subseteq L \times P \times L$ is a set of transitions.
\item $X$ is a set of variables.
\item For each transition $\tau \in T$: 
\begin{itemize}
\item $g_{\tau}$ is a Boolean condition over $X$: the guard of $\tau$,
\item $f_\tau = \{ (x, f^x(X)) \mid x\in X\}$ where $(x,f^x(X)) \in f_\tau$ expresses the assignment statement $x := f^x(X)$ updating $x$ with the value of the expression $f^x(X)$. 
\end{itemize}
\end{itemize}
\end{definition}
For $\tau = (l,p,l')\in T$ a transition of the internal LTS, $l$ (resp. $l'$) is referred to as the source (resp.
destination) location and $p$ is a port through which an interaction with another component can take place. Moreover, a transition $\tau = (l,p,l')\in T$ in the internal LTS involves a transition in the atomic component of the form $(l,p,g_\tau,f_\tau,l')$ which can be executed only if the guard $g_\tau$ evaluates to $\true$, and $f_\tau$ is a computation step: a set of assignments to local variables in $X$.

In the sequel we use the dot notation.
Given a transition $\tau = (l,p,g_\tau,f_\tau,l')$, $\tau.\source$, $\tau.\port$, $\tau.\guard$, $\tau.\func$, and $\tau.dest$ denote $l$, $p$, $g_\tau$, $f_\tau$, and $l'$, respectively.
Also, the set of variables used in a transition is defined as $\var(f_\tau) = \{x \in X \mid x:= f^x(X) \in f_\tau\}$.
Given an atomic component $B$, $B.P$ denotes the set of ports of the atomic component $B$, $B.L$ denotes its set of locations, etc.

Given a set $X$ of variables, we denote by $\valu{X}$ the set of valuations
defined on $X$. Formally, $\valu{X} = \{ \sigma : X \rightarrow \mathit{\Data}\}$, where
$\mathit{\Data}$ is the set of all values possibly taken by variables in $X$.

\paragraph{Semantics of Atomic Components.}
The semantics of an atomic component is an LTS over configurations and ports, formally defined as follows:
\begin{definition}[Semantics of Atomic Components]
The semantics of the atomic component $B = (P,L, T, X, \{g_{\tau}\}_{\tau \in T}, \{f_{\tau}\}_{\tau \in T})$ is defined as the labeled transition system  $S_B = (Q_B, P_B, T_B)$, where: 
\begin{itemize}
\item $Q_B = L\times \valu{X}$, where $\valu{X}$ denotes the set of valuations on $X$,
\item $P_B = P \times \valu{X}$ denotes the set of labels, that is, ports augmented with valuations of variables, 
\item $T_B$ is the set of transitions defined as follows. $T_B= \{ ((l',v'),p(v_{p}), (l,v))\in Q_B\times P_B\times Q_B \mid \exists \tau = (l', p[x_{p}], l) \in T: g_{\tau}(v') \wedge v=f_{\tau}(v'/v_{p})\}$, where $v_{p}$ is a valuation of the variables of $p$.  
\end{itemize}
\end{definition}
A configuration is a pair $(l,v)\in Q_B$ where $l \in L$ is a control location, $v \in \valu{X}$ is a valuation of the variables in $X$. The evolution of configurations $(l', v')\stackrel{p(v_{p})}{\rightarrow} (l, v)$, where $v_{p}$ is a valuation of the variables attached to the port $p$, is possible if there exists a transition $(l', p[x_p], g_\tau, f_\tau, l)$, such that $g_\tau(v')=\true$. As a result, the valuation $v'$ of variables is modified to $v=f_\tau(v'/v_p)$.
\subsubsection{Creating composite components.} Assuming some available atomic components $B_1,$ $\ldots,B_n$, we show how to connect the components in the set $\{B_i\}_{i\in I}$ with $I\subseteq [1,n]$ using an \emph{interaction}. An interaction $a$ is used to specify the sets of ports that have to be jointly executed.

\begin{definition}[Interaction]
\label{def:connector}
An interaction $a$ is a tuple $(P_a, G_a,F_a)$, where:
\begin{itemize}
\item $P_a \subseteq \cup_{i = 1} ^ { n } B_i.P$ is a nonempty set of ports that contains
at most one port of every component, that is, $\forall i: 1 \leq i \leq n: |B_i.P \cap P_a| \leq 1 $. We denote by $X_a = \cup_{p \in P_a} p.X$ the set of variables available to $a$,
\item $G_a : \valu{X_a} \rightarrow \{\true, \false\}$ is a guard, 
\item $F_a : \valu{X_a} \rightarrow \valu{X_a}$ is an update function. 
\end{itemize}
\end{definition}
$P_a$ is the set of connected ports called the support set of $a$. For each $i\in I$, $x_i$ is a set of variables associated with the port $p_i$.

\begin{definition}[Composite Component]
  A composite component is defined from a set of available atomic components $\{B_i\}_{i\in I}$ and a set of interactions $\gamma=\{a_j\}_{j\in J}$.
The connection of the components in $\{B_i\}_{i\in I}$ using the set $\gamma$ of connectors is denoted by $\gamma(\{B_{i}\}_{i\in I})$.
\end{definition}

\begin{definition}[Semantics of Composite Components]
\label{def-runtimesemanticscomposite}
A state $q$ of a composite component $\gamma(\{B_1, \ldots, B_n\})$, where $\gamma$ connects the $B_i$'s for $i\in [1,n]$, is an $n$-tuple $q=(q_1,\ldots,q_n)$ where $q_i=(l_i,v_i)$ is a state of $B_i$. Thus, the semantics of $\gamma(\{B_1, \ldots, B_n\})$ is precisely defined as the labeled  transition system $S = (Q,\gamma,\goesto)$, where:
\begin{itemize}
\item $Q= B_1.Q\times \ldots\times B_n.Q$, 
\item $\goesto$ is the least set of transitions satisfying the following rule:
\begin{mathpar}
\inferrule*
{
    a = (\{p_i\}_{i \in I},G_a,F_a) \in \gamma \and G_a(\{v_{p_i}\}_{i \in I}) \hva\\
    \forall i\in I:\ q_i \goesto[p_i(v_i)]_i q'_i \wedge v_i = F^i_{a}(\{v_{p_i}\}_{i \in I}) \and
    \forall i\not\in I:\ q_i = q'_i
}
{
    (q_1,\dots,q_n) \goesto[a] (q'_1,\dots,q'_n)
}
\end{mathpar}
where $v_{p_i}$ denotes the valuation of the variables attached to the port $p_i$ and $F^i_{a}$ is the partial function derived from $F_a$ restricted to the variables associated with $p_i$.
\end{itemize}
\end{definition}
The meaning of the above rule is the following: if there exists an interaction $a$ such that all its ports are enabled in the current state and its guard evaluates to \true, then the interaction can be fired. When $a$ is fired, all involved components evolve according to the interaction and uninvolved components remain in the same state. 

Notice that several distinct interactions can be enabled at the same time, thus introducing non-determinism in the product behavior.
One can add priorities to reduce non-determinism. In this case, one of the interactions with the highest priority is chosen non-deterministically.\footnote{The BIP engine implementing this semantics chooses one interaction at random, when faced with several enabled interactions.}
\begin{definition}[Priority]
  \label{defn:priority}
  Let $S = (Q,\gamma,\goesto)$ be the behavior of the composite component $\gamma(\{B_1, \ldots, B_n\})$.  A {\em priority model} $\pi$ is a
  strict partial order on the set of interactions $A$. Given a priority model $\pi$, we
  abbreviate $(a,a')\in \pi$ by $a \prec_\pi a'$ or $a \prec a'$ when clear from the context. Adding the priority model $\pi$ over $\gamma(\{B_1, \ldots, B_n\})$ defines a new composite component 
  $\pi\big(\gamma(\{B_1, \ldots, B_n\})\big)$ 
  noted $\pi(S)$ and whose behavior is defined by $(Q, \gamma, \goesto_\pi)$, where $\goesto_\pi$ is the least set of transitions satisfying the following rule:
\begin{mathpar}
\inferrule*
	{
      q \goesto[a] q' \and
      \neg\big(\exists a'\in A,\exists q''\in Q: a \prec a' \wedge q \goesto[a'] q'' \big)
    }
    {
      q \goesto[a]_\pi q'
    }
\end{mathpar}
\end{definition}
An interaction $a$ is enabled in $\pi(S)$ whenever $a$ is enabled in $S$ and $a$ is maximal according to $\pi$ among the active interactions in $S$.

Finally, we consider systems defined as a parallel composition of components together with an initial state.
\begin{definition}[System]
\label{def:system}
A BIP system ${\cal S}$ is a tuple $(B,\mathit{Init}, v)$ where $B$ is a composite component,  $\mathit{Init}\in B_1.L\times \ldots\times B_n.L$ is the initial state of $B$, and $v \in \valu{X^{Init}}$ where $X^{Init} \subseteq \cup_{i = 1} ^ { n } B_i.X$.
\end{definition}

Given a port $p$ from the system \Pm, we denote by (1) $interaction(p)$ to be the set of interactions that are connected to $p$; (2) $component(p)$ to be the component to which the port $p$ belongs; (3) $transitions(p)$ to be the set of transitions labeled by $p$. 

\section{One loop program (\caig) }
\label{sec:this}
A {\em one loop program} (\caig) ranges over boolean, 
integer and array variables. 
A variable can be either a register denoting a memory element or a wire denoting a functional macro.
An \caig starts with the variable declarations followed by the wire variable definitions. 
Then memory variables are initialized simultaneously using the \cci{do-together} construct. 
After initialization, an infinite loop keeps updating the value of the memory variables simultaneously. The listings in Figure \ref{fig:gr} shows the syntax of an \caig. 

The wires are defined in a list of assignment statements \cci{wiredef-list}. 
Each wire can be the target of at most one assignment statement. 
If a wire is not assigned then it is a non-deterministic {\em primary input}
with a new non-deterministic value at each iteration of the loop. 

The list of statements \cci{init-list} assigns initial values to 
the register variables. All the assignment statements within \cci{init-list} execute
simultaneously as indicated with the \cci{do-together} construct.

Similarly, the \cci{next-list} list of statements updates the values 
of the register variables. 

Each assignment statement has a left hand side target term 
which is either a variable or an access operator to an 
array element. 
The right hand side of an assignment is a combinational expression ranging over the program variables,  Boolean and arithmetic operators, and a ternary choice 
operator. The ternary choice \cci{(a? b : c)} returns $b$ if $a$ 
is \true and $c$ otherwise. 

\begin{figure}
\begin{tabular}{p{3cm}p{0.5cm}p{12cm}}
\begin{lstlisting}
decl-list

wiredef-list

do-together {
  init-list 
}

while(true) {
  do-together {
    next-list
  } 
}
\end{lstlisting}
&
&
\begin{lstlisting}
type: bool | int | bool [NUM] | int [NUM]; 
declaration: wire type id; | type id;

expr: term | uop expr| expr bop expr | expr ? expr : expr;
term: id | id[expr]; 

decl-list: declaration+
wiredef-list: (term = expr)*

init-list: (term = expr)* 
next-list: (term = expr)* 
\end{lstlisting}
\end{tabular}
\caption{\caig Syntax}
\label{fig:gr}
\end{figure}

\section{Sequential Circuit}
\label{sec:sequential}

The ABC synthesis and model checker reasons about And-Inverted-Graph 
representation of a sequential circuit.

\begin{definition}[Sequential circuit]
\rm A {\em sequential circuit} is a tuple $\big( (V, E),G,
O\big)$.  The pair $(V,E)$ represents a directed graph on
vertices $V$ and edges $E \subseteq V\times V$ where $E$
is a totally ordered relation.  The function $G: V \mapsto
{\mathit types}$ maps vertices to ${\mathit types}$.
There are three disjoint types: {\em primary inputs}, {\em
bit-registers} (which we often simply refer to as {\em
registers}), and logical {\em gates}.  Registers have designated
{\em initial values}, as well as {\em next-state
functions}.  Gates describe logical functions such as
the conjunction or disjunction of other vertices. 
A subset $O$ of $V$ is specified as the {\em
primary outputs} of $V$.  
We will denote the set of primary input variables by $I$,
and the set of bit-register variables by $R$.  
\label{def:back:seq_circuit}
\end{definition}

\begin{definition}[Fanins]
\rm We define the direct {\em fanin}s of a gate $u$ to be
$\{v \mid (v,u)\in E\}$ the set of source vertices connected
to $u$ in $E$.  We call the {\em support} of $u$ $\{v \mid
(v\in I \vee v \in R) \wedge (v,u) \in \ast E\}$ all
source vertices in $R$ or $I$ that are connected to $u$
with $\ast E$, the transitive closure of $E$.
\label{def:back:fanins} 
\end{definition}


For the sequential
circuit to be syntactically well-formed, vertices in $I$
should have no fanins, vertices in $R$ should have
2~fanins (the next-state function and the initial-value
function of that register), 
and every cycle in the sequential circuit should contain
at least one vertex from $R$.  The initial-value functions
of $R$ shall have no registers in their support.  All
sequential circuits we consider will be well-formed.  

The ABC analyzer reasons about And-Inverted-Graph (AIG)
sequential circuits which are
sequential circuits with only NAND gates restricted to have 2~fanins.
Since NAND is functionally complete, this is not a limitation.  

\subsection{Semantics of sequential circuits}
\label{s:back:crct_semantics}

\begin{definition}[State]
\rm A {\em state} is a Boolean valuation to vertices in $R$. 
\end{definition}

\begin{definition}[Trace]
\rm A {\em trace} is a mapping $t: V \times \mathbb{N} \rightarrow
\mathbb{B}$ that assigns a valuation to all vertices in
$V$ across time {\em steps} denoted as indexes from
$\mathbb{N}$.  The mapping must be consistent with $E$ and
$G$ as follows.  Term $u_{j}$ denotes the source vertex of
the $j$-th incoming edge to $v$, implying that
$(u_{j},v)\in E$.  The value of gate $v$ at time $i$ in
trace $t$ is denoted by $t(v,i)$.
\[
t(v,i)=
   \begin{cases}
      s^i_{v}            &:v \in I \ \text{with sampled value $s_{v}^i$}\\
      t(u_1, 0)       &:v \in R,i=0,u_1:=\ \text{initial-state of $v$}\\
      t(u_2, i-1)        &:v \in R,i>0,u_2:=\ \text{next-state of $v$}\\
      G_v\big(t(u_{1},i),...,t(u_{n},i)\big) &: v \ \text{is a combinational gate with function 
$G_v$}
   \end{cases} \newline
\]
\end{definition}

The semantics of a sequential circuit are defined with
respect to semantical traces.  Given an input valuation
sequence and an initial state, the resulting trace is a
sequence of Boolean valuations to all vertices in $V$
which is consistent with the Boolean functions of the
gates.  We will refer to the transition from one valuation
to the next as a {\em step}.  A node in the circuit is
justifiable if there is an input sequence which when
applied to an initial state will result in that node
taking value $\mbox{true}$.  A node in the circuit is
valid if its negation is not justifiable.  We will refer
to targets and invariants in the circuit; these are simply
vertices in the circuit whose justifiability and validity
is of interest respectively.
A sequential circuit can naturally
be associated with a finite state machine (FSM),
which is a graph on the states.  However, the 
circuit is very different from its FSM; among
other differences, it is exponentially more succinct in
almost all cases of interest~\cite{BuClMcDiHw92}. 


\subsection{Translation from \caig to AIG circuits }

The translation of an \caig into an AIG circuit
first constructs registers, wires, and primary input variables for each 
\caig variable. 
It then recursively traverses the right hand side 
of the assignment statements in \cci{wiredef-list}, \cci{init-list},
and \cci{next-list} to build corresponding combinational circuits.
It connects the outputs of the combinational 
circuits built for the \cci{init-list} and \cci{next-list} 
assignment expressions to the 
initial value and next state value input pins of the corresponding 
registers, respectively. 
Finally, it connects the outputs of the combinational circuits built 
for the  \cci{wiredef-list} to the wires referring to the 
variables declared as wire variables in \cci{decl-list}.

\paragraph{Variables.} 
We consider each variable not declared as a wire in \cci{decl-list}.
We instantiate a corresponding 
vector of AIG registers with an adequate bit width. 
The width of the bit vector can be selected by the user, 
or can be set to match the default width of the declared type. 
Typically the default values for the bit width are 
32 bits for an integer, one bit for a Boolean, and a 
finite two dimensional bit vector for an array. 
In our case, and for \caig programs generated from BIP systems, 
we will not have arrays of register variables
and we will only have fixed size arrays of Boolean wires as 
discussed in Section~\label{s:bip2caig}.
We consider variables declared as wires in \cci{decl-list}
and that do not have a corresponding assignment statement 
in \cci{wiredef-list} as non-deterministic. 
For each non-deterministic variable we instantiate a corresponding
vector of primary inputs with an adequate bit-width. 
We consider variables declared as wires in \cci{decl-list} with 
a corresponding assignment statement in \cci{wiredef-list} as functional macros. 
For each functional macro we 
instantiate a vector of identity gates (a sequence of two negation gates) 
where the fanouts correspond to the wire variable and the fanins correspond to
the expression defining the wire variable in \cci{wiredef-list}. 
We denote the gates corresponding to each variable by the function \cci{vargates}. 

\begin{lstlisting}
/*@\textbf{variables}@*/(decl-list, wiredef-list)
  foreach variable $v$ in decl-list
    if ($v$ is not a wire) 
      vargates($v$) = instantiate-registers($v$,type($v$))
    elseif ($v$ is not assigned in wiredef-list) 
      vargates($v$) = instantitate-primary-inputs($v$,type($v$))
    else 
      vargates($v$) = instantiate-identity-gates($v$,type($v$))
    endif
  endfor
\end{lstlisting}

\paragraph{Assignment statements.}
We consider each assignment statement in \cci{wiredef-list}, \cci{init-list},
and \cci{next-list} and traverse the right hand side expressions of
each assignment with the \cci{traverse} routine. 
The traversal of an expression runs recursively. 
If the expression refers to a variable $v$ (base case), 
then the traversal returns \cci{vargates(v)}. 
If the expression is a logical, conditional, or arithmetic expression, then
the \cci{library} routine looks it up in a complete table of circuits
with the adequate bit width. 
For example, if the expression is a ternary conditional statement of the
form $b?~e_1:e_2$, then \cci{library} instantiates a multiplexer, 
connects its two data fanins to the nodes corresponding to $e_1$ and $e_2$, 
connects its control fanins to the nodes corresponding to $b$,
and returns its fanouts. 

\begin{lstlisting}
/*@\textbf{traverse}@*/($exp$)

  if ($exp$ is a variable) 
    return vargates(exp)
  endif

  foreach $i [1~..~ exp.operands.size()]$ 
    $wirevec[i]$ = traverse($exp.operands[i]$) 
  endfor

  return library($exp.operation$, $wirevec$)
\end{lstlisting}

\paragraph{Connections.}
Finally, we connect the nodes corresponding to the right hand side expressions 
of the assignment statements in the \cci{init-list} and \cci{next-list}
expressions 
to the initial value and next value fanins of the corresponding register gates, 
respectively. 
We connect the nodes corresponding to the right hand side expressions
of the assignment statements in the \cci{wiredef-list} expressions to the 
fanins of the corresponding wire identity gates.

\section{BIP to \caig}
\label{sec:bip2aig}

Given a BIP system $\Pm = (B, Init, v)$, \biptool~
calls function  \cci{BIP-to-OLP} to translate \Pm into 
an \caig~program with its own customized execution engine. 
It calls four functions 
that fill \cci{decl-list}, \cci{wiredef-list}, \cci{init-list}, \cci{next-list}. 
All these function use the \cci{append} call to add code fragments to lists. 

\begin{lstlisting}
/*@\textbf{BIP-to-OLP}@*/(B, Init, v)
  generateDeclartionList()
  generateWireDefList()
  generateInitList()
  generateNextList()
\end{lstlisting}

Function \cci{generateDeclartionList()} fills \cci{decl-list} as follows. 
It creates three arrays of wires to denote interaction semantics. 
Array $ie$ elements denote whether all logical constraints except priority rules are met for a given interaction. 
Array $ip$ elements denote whether a given interaction is enabled after applying priority rules. 
Array $is$ elements denote whether an enabled interaction is selected for execution. 
Currently, one interaction is selected to avoid executing conflicting interactions. 
Two interactions are conflicting if they involve same components. 
The $selector$ wire is a non-deterministic primary input that is used to select one of the enabled interactions. 
The $cycle$ boolean register is used to denote whether the system is executing actions corresponding to either interaction or transition. 
The function also declares two wires ($B_i.p_j.e$ and $B_i.p_j.s$) for each port $p_j$. 
Wire $B_i.p_j.e$ denotes whether the port is enabled and wire $B_i.p_j.s$ denotes whether the port is selected by the interaction for execution. 
Moreover, for each component $B_i$ the function declares a register variable $B_i.\ell$ denoting the current location of $B_i$. 
Similarly, the function declares a variable register $B_i.x_j$ for each variable $x_j$ in component $B_i$.  

\begin{lstlisting}
/*@\textbf{generateDeclartionList()}@*/
  // interaction enablement wires
  append $wire~bool~ie[|J|]$ to decl-list
  // interaction priority wires
  append $wire~bool~ip [ |J| ]$ to decl-list 
  // interaction selected wires
  append $wire~bool~is[|J|]$ to decl-list 
  // non-deterministic priority selector wire
  append $wire~int~selector$ to decl-list 
  // cycle denotes transition or interaction mode
  append $bool~cycle$ to decl-list  

  foreach $i \in [1..|I|]$
    foreach $j \in [1..|B_i.P|]$ 
      // port enablement
      append $wire~bool~B_i.p_j.e$ to  decl-list 
      // port selected
      append $wire~bool~B_i.p_j.s$ to  decl-list 
    endfor

    // location registers
    append $int~B_i.\ell$ to decl-list
    
    foreach $j \in [1..|B_i.X|]$ 
      // variable registers
      append $int~B_i.x_j$ to  decl-list 
    endfor
  endfor
\end{lstlisting}

Function \cci{generateWireDefList()} fills \cci{wiredef-list} with functional macro definitions as follows. The enable wire $B_i.p_j.e$ is \true when there exists a transition $\tau$ labeled with port $p$, its source ($\tau.src$) is the current location ($B_i.\ell$), and its guard is \true. 

Array element $ie[j]$ corresponding to interaction $a_j$ is evaluated to \true when the guard of $a_j$ is \true and all its ports are enabled. Array element $ip[j]$ is evaluated to \true when $ie[j]$ is \true and $a_j$ has higher priority than other enabled interactions. Array element $is[j]$ is evaluated to \true when $ip[j]$ is true either $a_j$ is selected ($selector$ equals to $j$) or the selected interaction is not enabled and $j$ is the first enabled interaction greater with an index greater than $j$. The use of non-deterministic selector is added for fairness. 
The selected wire $B_i.p_j.s$ is \true when there exists a selected interaction $a_k$ (i.e., $is[k]$ is \true) involving $B_i.p_j$.

\begin{lstlisting}
/*@\textbf{generateWireDefList()}@*/
  // iterate over components
  foreach $i \in [1..|I|]$ 
    // iterate over component ports
    foreach $j \in [1..|B_i.P|]$ 
      append $B_i.p_j.e := \bigvee_{\tau \in transitions(B_i.p_j)} \tau.guard \land B_i.\ell = \tau.src$ to  wiredef-list 
    endfor
  endfor
  
  // iterate over interactions
  foreach $j \in [1..|J|]$ 
    append $ie[j] := a_j.guard \wedge \bigwedge_{p\in a_i.P} component(p).p.e$ to  wiredef-list 
    append $ip[j] := ie[j] \wedge \,(\forall k \neq j: ie[k] \Rightarrow a_k < a_j)$ to  wiredef-list 
    append $is[j] := ip[j] \wedge \,(selector = j \vee (\lnot ip[selector] \land \forall k > j: \neg ip[k])$ to  wiredef-list 
  endfor
  
  // iterate over components
  foreach $i \in [1..|I|]$ 
    // iterate over component ports
    foreach $j \in [1..|B_i.P|]$ 
      append $B_i.p_j.s := \bigvee_{a_k \in interactions(B_i.p_j)} is[k]$ to  wiredef-list 
    endfor
  endfor
\end{lstlisting}

Function \cci{generateInitList()} fills \cci{init-list} with initial value definitions taken from $Init$ for location variables ($B_i.\ell)$ and $v$ for component variables ($B_i.x_j)$. Register variable $cycle$ is initialized to zero to denote an interaction execution mode. 

\begin{lstlisting}
/*@\textbf{generateInitList()}@*/
  // initialize to interaction mode
  append $cycle := 0$ to init-list 
  foreach $i \in [1..|I|]$
    append $B_i.\ell := Init.B_i$ to  init-list 
    foreach $j \in [1..|B_i.X|]$
      // v is the initial valuation
      append $B_i.x_j := v(B_i.x_j)$ to init-list  
    endfor
  endfor
\end{lstlisting}

Function \cci{generateNextList()} fills \cci{next-list} with the next state value definitions of register variables. Each component variable can be modified either in an interaction action or in a transition action. The value of variable $cycle$ makes this distinction. 

In the interaction mode ($cycle$ equals to zero), the function considers each assignment statement $\sigma$ from the action of interaction $a_j$. The function appends a conditional clause requiring the $a_k$ to be selected for execution so that the target variable $B_i.x_j$ of $\sigma$ is assigned to the expression of $\sigma$ ($\sigma.expr$). The sequence of conditional clauses form a nested ternary conditional expressions where the last expression retains the previous value of the variable. 

Similarly, in the transition mode ($cycle$ equals to one), the function considers each assignment statement $\sigma$ from the action of transition $\tau$. The function appends a conditional clause requiring the port of the transition $\tau$ to be selected for execution and the location of the component to be equal to the source of the transition. The target variable $B_i.x_j$ of $\sigma$ is assigned to the expression of $\sigma$ ($\sigma.expr$). 

In the transition mode, the function considers the current location of each component $B_i.\ell$ and appends a conditional clause requiring the transition source to be equal to the current location and the port of the transition to be selected. The expression corresponding to the conditional clause updates the current location to be the destination of the transition ($\tau.dest$).  In the interaction mode, the location retains its value. Finally, the $cycle$ variable is toggled. 

\begin{lstlisting}
/*@\textbf{generateNextList()}@*/ 
  // iterate over components - interaction-mode
  foreach $i \in [1..|I|]$ 
    // iterate over variables, where $\textcolor{darkgreen}{B_i.X = \{x_1, \ldots, l_{|B_i.X|}\}}$ 
    foreach $j \in [1..|B_i.X|]$ 
      // interaction mode
      append $B_i.x_j := cycle = 0?$ to var-st
      // iterate over interactions
      foreach $k \in [1..|J|]$ 
        // iterate over interaction assignments
        foreach $\sigma \in a_k.action$
          if($B_i.x_j = \sigma.term$)
            append $is[k]?\, \sigma.expr:$ to var-st
          endif
        endfor
      endfor
      // interaction mode and no data transfer for $\textcolor{darkgreen}{B_i.x_j}$
      append $B_i.x_j$: to var-st 
      
      
      // iterate over component transitions - transition-mode
      append $B_i.\ell := cycle = 0?\, B_i.\ell :$ to loc-st
      foreach $\tau \in B_i.T$ 
        // iterate over transition assignments
        foreach $\sigma \in \tau.action$
          if($B_i.x_j = \sigma.term$)
            append $(B_i.port(\tau).s \wedge \tau.src = B_i.\ell)?\, \sigma.expr:$ to var-st 
          endif
        endfor
        append $(B_i.port(\tau).s \wedge \tau.src = B_i.\ell)?\, \tau.dest:$ to loc-st 
      endfor  
      
      append $B_i.x_j$ to var-st 
      append var-st to  next-list 
      
      append $B_i.\ell$ to loc-st 
      append loc-st to  next-list 

    endfor
    // switch cycle
    append $cycle := \neg cycle$ to  next-list 
  endfor
\end{lstlisting}

\subsection{Illustrative Example}
\begin{figure}[h!]
 \centering
 \resizebox{0.5\textwidth}{!}{
   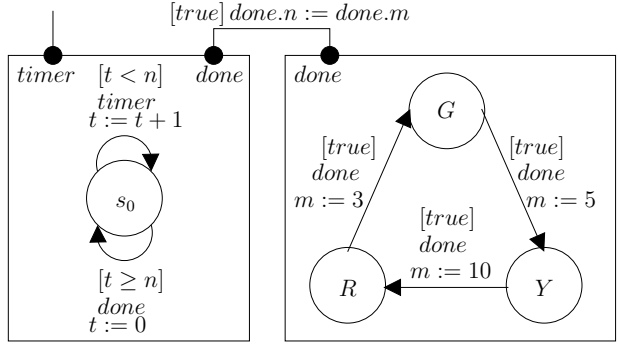
 }
 \caption{Traffic light in BIP}
 \label{fig:traffic:bip}
\end{figure}

Figure \ref{fig:traffic:bip} shows a traffic light controller system modeled in BIP. 
It is composed of two atomic components, \cci{timer} and \cci{light}. The timer counts
the amount of time for which the light must stay in a specific state (i.e. a specific
color of the light). The light component determines the color of the traffic light. Additionally, it
informs the timer about the amount of time to spend in each location through a data transfer on the interaction between the two components. 

Figure~\ref{fig:caigtraffic} shows $\caig$ code generated that corresponds to the  to the traffic light controller BIP system shown in Figure~\ref{fig:traffic:bip}. 

\begin{figure}
\begin{tabular}{p{4.2cm}p{.2cm}p{12.5cm}}
\begin{lstlisting}
/*** decl-List ***/
int timer.t;
int timer.n;
int light.m;
int timer.$\ell$;
int light.$\ell$;
bool cycle;

wire int selector;
wire bool timer.timer.e;
wire bool timer.timer.s;
wire bool timer.done.e;
wire bool timer.done.s;
wire bool light.done.e;
wire bool light.done.s;
wire bool ie[2];
wire bool ip[2];
wire bool is[2];
\end{lstlisting}
& & 
\begin{lstlisting}
/*** wiredef-list ***/
timer.timer.e = (0 == timer.$\ell$) && (timer.t < timer.n);

timer.done.e  = (0 == timer.$\ell$) && (timer.t == timer.n);
light.done.e  = (0 == light.$\ell$)  || (1 == light.$\ell$) || (2 == light.$\ell$);

ie[0] = timer.timer.e;
ie[1] = (light.done.e && timer.done.e);

ip[0] = ie[0];
ip[1] = ie[1];

is[0] = (ip[0] && ( selector == 0 || (!ip[selector]  && !ip[1]);
is[1] = (ip[1] && ( selector == 1 || (!ip[selector]);

timer.timer.e = is[0];
timer.done.e  = is[1];
light.done.e  = is[1] ;
\end{lstlisting}
\\
\vspace{-2em}
\begin{lstlisting}
do-together {
  /*** init-list ***/
  timer.t = 0; 
  timer.n = 10; 
  timer.$\ell$ = 0;

  light.m = 5; 
  light.$\ell$ = 0;

  cycle = true; 
}/* end do-together */
\end{lstlisting}
& & 
\vspace{-2em}
\begin{lstlisting}
while(true) {
  do-together {
    /*** next-list ***/    
    timer.n = cycle? is[1]? light.m : timer.n : timer.n; 
    
    timer.$\ell$ = cycle? timer.$\ell$: timer.timer.e && timer.$\ell$ == 0? 0 : timer.timer.e && timer.$\ell$ == 0? 0 : timer.$\ell$;
    
    timer.t = cycle? timer.t : timer.$\ell$ == 0 && timer.timer.e? (timer.t + 1) : timer.$\ell$ == 0 && timer.done.e? 0 : 
              timer.t; 
              
    light.$\ell$ = cycle? light.$\ell$ : light.$\ell$ == 2 && light.done.e? 0: light.$\ell$ == 1 && light.done.e? 0 : light.$\ell$ == 0 
              && light.done.e? 1 : light.$\ell$; 
              
    light.m = cycle? light.m : light.$\ell$ == 0 && light.done.e? 3: light.$\ell$ == 1 && light.done.e? 10: light.$\ell$ == 2 
              && light.done.e? 5 : light.m; 
    
    cycle = !cycle; 
  } /*end do-together*/ 
} /*end while(true)*/
\end{lstlisting}
\end{tabular}
\vspace{-2em}
\caption{Sample of $\caig$ generated code of traffic light system}
\label{fig:caigtraffic}
\end{figure}

\subsection{One cycle optimization}
Recall that an interaction specifies a strong synchronization among its involved components. Data transfer can take place during such synchronization. The operational semantic of BIP requires to (1) first execute data transfer of the selected interaction and then (2) execute the functions of the corresponding transitions of atomic components. For this, in the above translation, we used $cycle$ boolean register to denote whether the system is executing actions corresponding to either interaction or transition. However, in some cases data transfers of all interactions modify some variables that are not assigned in the corresponding transitions of those interactions. This can be detected by doing a static data dependency between interactions and their transitions. 
This may drastically improve the performance of the system since data transfers as well as functions of transitions may be executed in one cycle. Note that, our implementation supports this optimization.

\section{Implementation and Evaluation}
\label{sec:implem}

\subsection{\biptool{}} \label{chap:implementation:bip}
\biptool{} is a Java implementation of the translation from BIP to $\caig$ described in Section \ref{sec:bip2aig}, and, is part of the BIP distribution~\cite{verimagbiponline}.

\biptool~ takes as input a BIP system and a set of invariants
and generates the corresponding \caig{} program with a system-specific 
execution framework. 
The \caig program can be directly compiled for runtime verification (simulation)
where primary inputs are set to random values at each iteration. 
The resulting binary can be concurrent by replacing the 
\cci{do-together} construct with the corresponding openmp directives. 

For sequential synthesis, we synthesize an AIG 
circuit that can be used as an FPGA implementation of the system
and pass it to ABC. 
We use ABC synthesis and reduction algorithms to reduce the area and the critical
time of the AIG circuit by 
removing latches and logic gates using techniques such as 
retiming~\cite{KuBa01}, 
redundancy removal~\cite{HmBPK05,KuMP01,BjesseC00,aziz-fmsd-00}, 
logic rewriting~\cite{BjBo04}, interpolation~\cite{McMillan03}, 
and localization~\cite{Wang03}. 
The reduced AIG circuit is equivalent to the original circuit and can be used
as a reduced FPGA implementation. 


For verification, 
ABC uses the sequential synthesis techniques above to reduce the 
AIG circuit and render it amenable for decision algorithms. 
Then ABC uses decision algorithms such as 
symbolic model checking, bounded model checking, induction, 
interpolation, circuit SAT solving, 
and target enlargement~\cite{MoGS00,MoMZ01,HoSH00,BaKuAb02,Hari05expert}
to verify the correctness of the circuit with respect to the BIP system invariants.
It either proves correctness or produces a counter example where the system 
violates the property. 
\biptool~ provides a debugging mechanism where the counter example is mapped back 
to the original BIP system. 
The debugging tool is integrated with wave form visualization 
tool \cite{bybell2010gtkwave}.  

\biptool~ is equipped with a command line interface that accepts a set 
of configuration options. 
It takes the name of the input BIP file and optional flags. 
\begin{lstlisting}[language=Bash]
java -jar bip-to-abc.jar [options] input.bip output.abc [property.txt]
\end{lstlisting}

We evaluated \biptool{} against two industrial benchmarks, 
an {\em Automatic Teller Machine} (ATM)~\cite{atm} and the {\em Quorum} consensus
protocol~\cite{guerraoui2012speculative}. We report on the size of the generated
AIGs before and after reduction, and on the time taken by the ABC solver to 
reduce and verify the benchmarks. We compare the results for the 
verification of the ATM benchmark with the NuSMV~\cite{nusmv} model checker.

\subsection{The ATM benchmark}
\begin{figure}[bt]
 \centering
  \resizebox{1.0\textwidth}{!}{
       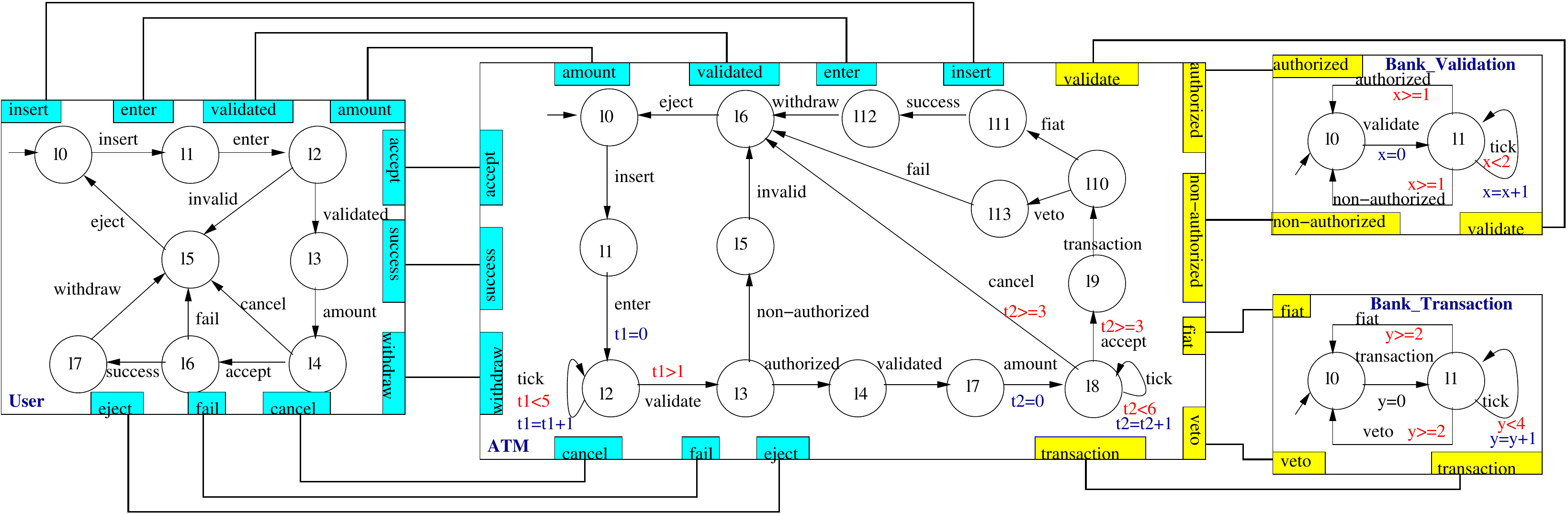
  }
  \caption{Modeling of ATM system in BIP} 
  \label{fig:atm_bip}
\end{figure}

Automatic Teller Machine (ATM) is a computerized system that provides financial services for users in 
a public space. Figure~\ref{fig:atm_bip} shows a structured BIP model of an ATM system adapted from the
description provided in~\cite{atm}.
The system is composed of four atomic components: (1) the User (2) the ATM (3) the Bank Validation 
and (4) the Bank Transaction. ATM component handles all interactions between the 
users and the bank. No communication between the users and the bank is allowed. 

The ATM starts from an idle location and waits for the user to insert the card 
and enter the confidential code. The user has $5$ time units
to enter the code before the counter expires and the card is ejected by the ATM. 
Once the code is entered, the ATM checks with the bank validation unit for 
the correctness of the code. If the code is invalid, the card is ejected
and no transaction occurs. If the code is valid, the ATM waits for the user to enter
the desired amount of money for the transaction. The time-out for entering the amount 
of money is of $6$ time units. 

Once the user enters the desired transaction amount, the ATM checks with the bank whether 
the transaction is allowed or not by communicating with the bank transaction unit.
If the transaction is approved, the money is transferred to the user and the card is ejected. 
If the transaction is rejected, the user is notified and the card is ejected. In all cases, 
the ATM goes back to the idle location waiting for any additional users. 
In our model, we assume the presence of a single bank and multiple ATMs and users. 

\begin{table}[tb]
\centering
\begin{tabular}{|c|c|c|c||c|c|c||c|c|}
\cline {2-9}
\multicolumn{1}{c|}{} &  \multicolumn{3}{c||}{Original} & \multicolumn{3}{c||}{After reduction} &  \multicolumn{2}{c|}{Time(s)} \\ \hline
ATMs & latches & AND-gates & levels & latches & AND-gates & levels & \biptool& NuSMV \\ \hline
2 & 78 & 2308 & 125 & 37 & 552 & 25 & 26.1 & 1.4\\ \hline
3 & 102 & 3689 & 197 & 50 & 804 & 29 & 32.65 & 142.6 \\ \hline
4 & 146 & 5669 & 234 & 63 & 1036 & 29 &  597 & 3361 \\ \hline
\end{tabular}
\caption{ATM results}
\label{tb:bip:atm}
\end{table}

Table~\ref{tb:bip:atm} shows the improvement obtained by using \biptool{}
to verify the deadlock freedom of the ATM system, as compared to using the
NuSMV model checker~\cite{nusmv}.
The first column of the table shows the number of clients and ATMs in the system. 
The table contains the number  of latches, AND gates and logic levels in the AIG generated by \biptool{} before and after applying reduction techniques, respectively.
We report on the verification time taken by the ABC solver to check the 
generated AIG, and the total taken to perform both synthesis (reduction) 
and verification, in addition to the time taken by NuSMV to perform verification.

With the increase in the number of users and ATMs in the system, \biptool{}  
outperforms NuSMV in terms of total verification time, reaching a speedup 
of $5.6$ for 4 users and ATMs. Additionally, \biptool{} allows developers
to make use of several reduction techniques that are able to reach an 
average of $50\%$ reduction in the size of the AIG. Note that for $2$ ATMs 
and users, NuSMV outperforms \biptool{}. This is due to the fact that when 
performing verification, ABC tries multiple verification and reduction 
algorithms before reaching a conclusive result. However, the advantage 
that \biptool{} presents can be clearly seen for larger number of ATMs and 
users.

\subsection{The Quorum protocol}
The {\em Quorum} protocol is a consensus protocol proposed in~\cite{guerraoui2012speculative}
as complementary to the Paxos consensus protocol~\cite{gafni2003disk} under perfect
channel conditions. {\em Consensus} allows a set of communicating processes
(clients and servers in our case) to agree on a common value. Each of clients proposes
a value and receives a common decision value. The authors in~\cite{guerraoui2012speculative}
propose to use Quorum when no failures occur (perfect channel conditions) and 
Paxos when less than half of the servers may fail. 

The Quorum protocol operates as follows.
\begin{enumerate}
 \item Upon proposal, a client $c$ broadcasts its proposed value 
 $v$ to all servers. It also saves $v$ in its local memory and starts a local time
 $t_c$. 
 \item When a server receives a value $v$ from a client $c$, it performs
 the following check.
 \begin{itemize}
  \item It if has not sent any accept messages, it sends an accept message
  $accept(v)$ to the client $c$. 
  \item If it has already accepted value $v'$, it sends an accept message
  $accept(v')$ to the client $c$. 
 \end{itemize}
 \item If a client $c$ receives two different accept messages, it switches
 to the backup phase $switch-backup(proposal_c)$.
 \item If a client $c$ receives the same accept messages $accept(v)$ from all the servers,
 it decides on the value $v$.
 \item If a client's timer $t_c$ expires, it waits for at least
 one accept message $accept(v')$ from a server, or chooses a value $v'$
 from an already received $accept(v')$ message, and then switches to 
 the backup phase with the value $v'$. 
 \item The {\em backup} phase is an implementation of the Paxos algorithm. Quorum in this
 case has decided that the channel is not perfect. 
\end{enumerate}

We implemented the Quorum protocol in BIP, and we used \biptool{} to verify 
two invariants as defined in~\cite{guerraoui2012speculative}.
\begin{enumerate}
 \item $Invariant_1$: If a client $c$ decides on a value $v$, then all clients 
 $c' \neq c$ that have switched, either before or after $c$, switch with the value $v$.
 \item $Invariant_2$: If a client $c$ decides on a value $v$, then all clients
 $c' \neq c$ who decide, do so with the same value $v$. 
\end{enumerate}

Table~\ref{tb:bip:qrm} shows the results of using \biptool{} to verify the 
Quorum protocol for $2$ and $4$ clients with $2$ servers. The designs
are indexed as \cci{num\_clients}-\cci{num\_servers}-\cci{status} where 
\cci{num\_clients} is the number of clients, \cci{num\_servers} is the number of 
servers and \cci{status} is either valid (\cci{v}) or erroneous (\cci{e}).
A valid design contains no design bugs, while an erroneous design is injected
with a bug. We report on the size of the AIG in terms of number of latches,
number of AND gates and logic levels before and after
applying reduction algorithms.

\begin{table}[bt]
\centering
\begin{tabular}{|c|c|c|c||c|c|c||c|c|}
\cline{2-9}
\multicolumn{1}{c|}{} & \multicolumn{ 3}{c||}{Original} & \multicolumn{3}{c||}{After reduction} & \multicolumn{ 2}{c|}{Time (s)} \\ \hline
Design & latches & AND-gates & levels & latches & AND-gates & levels & \biptool& NuSMV \\ \hline
2-2-e & 264 & 3508 & 101 & 65 & 923 & 51 & 0.78 & 526 \\ \hline
2-2-v & 264 & 3614 & 105 & 66 & 641 & 29 & 240.6 & 526 \\ \hline
4-2-e & 390 & 6305 & 145 & 117 & 1129 & 50 & 0.24  & memory-out \\ \hline
4-2-v & 390 & 6453 & 151 & 117 & 1170 & 30 & 58 hours & memory-out \\ \hline
\end{tabular}
\caption{Quorum results}
\label{tb:bip:qrm}
\end{table}

\begin{figure}[bt]
\centering
\scalebox{0.75}{
 \includegraphics{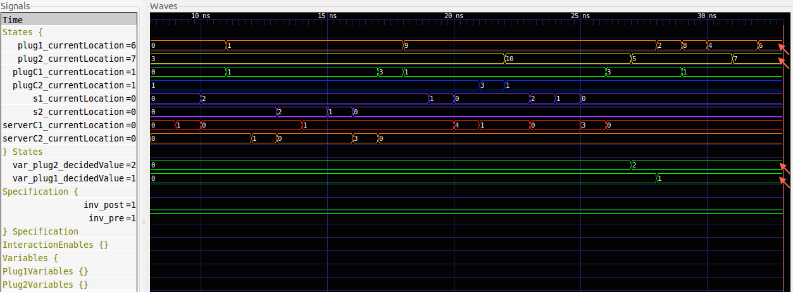}}
\caption{Visualization of a counter example using Gtkwave}
\label{fig:res:counter}
\end{figure}

Using ABC's synthesis and reduction algorithms, \biptool{} was able to
reduce the size of the generated AIGs for all designs by a factor larger
than $50\%$. Furthermore,
\biptool{} was able to give conclusive results about all four designs, unlike
NuSMV which failed to give any decision about the designs having
$4$ clients and $2$ servers. For example, \biptool{} found a counter example for the erroneous 
design having $4$ clients and $2$ servers in $0.24$ sec while NuSMV failed to do
so. Figure~\ref{fig:res:counter} shows a snippet of the generated counter example for the 
erroneous design, visualized using the Gtkwave~\cite{bybell2010gtkwave} waveform viewer. 
The variables presented in the counterexample are the current control locations and the value of the variables of the different components in the design. Red arrows points to the values that implies a violation of the invariant. 

\section{Related work}
\label{sec:related}

The overlap between software and hardware design in embedded systems creates more challenges 
for verification and code generation. 

SystemC~\cite{systemc} is a modeling platform based on C++ that provides
design abstractions at the {\em Register Transfer Level} (RTL), behavior, and system levels. 
It aims at providing a common design environment for embedded system design and hardware-software
co-design. SystemC designers write their systems in C++ using SystemC class libraries that 
provide implementations for hardware specific objects such as concurrent modules, synchronization constructs,
and clocks.
Therefore the input systems can be compiled using standard C++ compilers to generate binaries
for simulation. SystemC allows for the communication between different components of a system
through the usage of ports, interfaces and channels.  

Metropolis~\cite{metropolis1,metropolis2} is an embedded system design platform based on 
formal modeling and separation of concerns for an effective
design process.
A Metropolis process is a sequence of events representing
functionality, and different processes communicate via ports of interfaces.
An interface includes methods that processes can use to communicate. 
Metropolis uses SIS for synthesis, SystemC and Ptolemey for runtime verification, and SPIN for model checking.
While BIP separates behavior from interaction (synchronization and communication) to simplify correctness by construction
and compositional verification, Metropolis separates communication from behavior (computation) and leaves synchronization 
highly coupled within each of them.

The BIP framework differs from SystemC in that it presents a dedicated language and supporting
tool-set that describes the behavior of individual system components as symbolic LTS. 
Communication between components in BIP is ensured through ports and interactions.   
BIP operates at a higher level than SystemC and does not provide support for circuit level 
constructs.

Verification techniques for SystemC and BIP make use of symbolic model checking tools. 
NuSMV2~\cite{nusmv} is a symbolic model checker that employs both 
SAT and BDD based model checking techniques. It processes an input 
describing the logical system design as a finite state machine, and a set of specifications
expressed in LTL, Computational Tree Logic (CTL) and Property Specification Language (PSL).
Given a system $\Pm$ and a set of specifications $P$, NuSMV2 first flattens $\Pm$ and $P$ by 
resolving all module instantiations and creating modules and processes, thus generating one 
synchronous design. It then performs a Boolean encoding step to eliminate all scalar variables, 
arithmetic and set operations and thus encode them as Boolean functions.   

The work in \cite{NiakiDATAS13} uses constraint based
programming to compute an executable 
MPI based parallel simulator of an embedded and cyber-physical 
system written in ForSyDe~\cite{SanderJ04}.
ForSyDe is a library of SystemC based 
parametrized system components with strict constraint 
specifications and a blocking write FIFO queue modeling 
a Kan network. 
The instants of the ForSyDe components are processes that 
communicate only through signals. 

The work in \cite{BarnatVMCAI2013} introduces a model checking
methodology for LTL specifications of embedded system written 
in DIVINE~\cite{Divine}  over
a total store order (TSO) of memory elements. 
Our method assumes a similarly relaxed memory model
since it adopts a cycle based execution model where 
updated memory values are observable at the next cycle. 

In order to avoid the state space explosion problem, NuSMV2 performs a cone of 
influence reduction~\cite{berezin1998compositional} step in order to eliminate
non-needed parts of the flattened model and specifications. The cone of influence
reduction abstraction technique aims at simplifying the model in hand by only 
referring to variables that are of interest to the verification procedure, i.e. variables
that influence the specifications to check~\cite{clarke1999model}.

DFinder~\cite{dfinder} is an automated verification tool for checking invariants
on systems described in the BIP language. Given a BIP system \Pm and 
an invariant $\mathcal{I}$, DFinder operates  compositionally and iteratively
to compute invariants $\mathcal{X}$ of the interactions and the atomic 
components of \Pm. It then uses the Yices {\em Satisfiability Modulo
Theory} (SMT) solver~\cite{dutertre2006fast} to check for the validity 
of the formula $\mathcal{X} \land \lnot \mathcal{I} = false$. 
Additionally, DFinder checks the deadlock freedom of  \Pm by building an invariant 
$\mathcal{I}_d$ that represents the states of of \Pm in which no interactions 
are enabled, \ie{} a deadlock occurs. It then checks the for the formula
$\mathcal{X} \land \mathcal{I}_d = false$, \ie{} none of the deadlock states
are reachable in \Pm.   

Techniques based on symbolic model checking for the verification of 
BIP designs suffer from the state space explosion problem, and often 
fail to scale with the size and the complexity of the systems. 
On the other hand, DFinder does not handle data transfer between 
atomic components, thus limiting the range of practical applications 
on which it can be applied. 
Our technique handles data transfers and uses the wide range of synthesis 
and reduction algorithms provided by ABC to effectively reduce the size and 
the complexity of the verification problem. Most of these algorithms have no counterpart
in symbolic model checking.  

Unlike all the methods described above, our method leverages
the same semantics for FPGA synthesis, model checking, 
and runtime verification (simulation). 

\section{Conclusion and Future Work}
\label{sec:conclusion}

In this paper we present a method for embedded system synthesis, runtime verification,
and model checking with supporting tools for the BIP framework. 
The method takes a BIP system and generates a concurrent C program with a system 
specific scheduler embedded therein. 
The concurrent C program serves as a software runtime verification simulator for the 
BIP system.
The method then take the concurrent C program and generates an AIG circuit which is an
FPGA implementation of the BIP system. 
The method applies synthesis reduction techniques using the ABC framework 
to simplify and reduce the AIG circuit
into a smaller and a less complex circuit that can be readily implemented with an 
FPGA. 
The method passes the reduced AIG circuit with a designated output that is \true
when the BIP system invariants are \true to ABC proof and model checking 
algorithms. In case ABC finds a counterexample, the methods maps the values from 
the counterexample to the original ABC system and provides the user with a debug
visualization tool. 
We successfully used the system to verify and debug medium and large case studies. 


Currently, the system specific scheduler makes conservative decisions to avoid interaction conflicts. Two interactions are conflicting if they share a port or they are using conflicting ports of the same component. An important extension is to allow parallel execution of non-conflicting interactions using techniques presented in \cite{BonakdarpourBJQS12}. 

\bibliographystyle{plain}
\bibliography{biblio}

\end{document}